\documentclass[sigconf]{acmart}

\usepackage{booktabs} % For formal tables

\copyrightyear{2020}
\acmYear{2020}
\setcopyright{acmcopyright}
\acmConference[JCDL '20]{Proceedings of the ACM/IEEE Joint Conference on Digital Libraries in 2020}{August 1--5, 2020}{Virtual Event, China}
\acmBooktitle{Proceedings of the ACM/IEEE Joint Conference on Digital Libraries in 2020 (JCDL '20), August 1--5, 2020, Virtual Event, China}
\acmPrice{15.00}
\acmDOI{10.1145/3383583.3398531}
\acmISBN{978-1-4503-7585-6/20/06}
\usepackage{todo}

\renewcommand\footnotetextcopyrightpermission[1]{}
\begin{document}
\fancyhead{}
\title{Unsupervised Anomaly Detection in Journal-Level \\Citation Networks}

\author{Baani Leen Kaur Jolly*, Lavina Jain*, Debajyoti Bera, Tanmoy Chakraborty}
\authornote{These authors contributed equally to this research.}
\affiliation{%
  \institution{IIIT-Delhi, India}
}
\email{{baani16234, lavina16052, dbera, tanmoy}@iiitd.ac.in}

\begin{abstract}
{Journal Impact Factor} is a popular metric for determining the quality of a journal in academia.
The number of citations received by a journal is a crucial factor in determining the impact factor, which may be misused in multiple ways. Therefore, it is crucial to detect citation anomalies for further identifying manipulation and inflation of impact factor.
{Citation network} models the citation relationship between journals in terms of a directed  graph.  
Detecting anomalies in the citation network is a challenging task which has several applications in spotting citation cartels and citation stack, and understanding the intentions behind the citations.

In this paper, we present a novel approach to detect the anomalies in a journal-level scientific citation network, and compare the results with the existing graph anomaly detection algorithms. 
Due to the lack of proper ground-truth, we introduce a journal-level citation anomaly dataset which consists of synthetically injected citation anomalies and use it to evaluate our methodology.
Our method is able to predict the anomalous citation pairs with a precision of 100\% and an F1-score of 86\%. 
We further categorize the detected anomalies into various types and reason out possible causes. We also analyze our model on the Microsoft Academic Search dataset - a real-world citation dataset and interpret our results using a case study, wherein our results resemble the citations and SCImago Journal Rank (SJR) rating-change charts, thus indicating the usefulness of our method.
We further design `Journal Citation Analysis Tool', an interactive web portal which, given the citation network as an input, shows the journal-level anomalous citation patterns  and helps users analyze citation patterns of a given journal over the years. 
\end{abstract}

\if 0


\fi

\maketitle

\section{Introduction}\label{introduction}
A citation network refers to a (un)weighted, directed graph with the directed edges representing the citations, the edge weights (if exist) representing the number of citations between two nodes (e.g., papers, journals, authors). The impact made by the work done by an entity/node is measured using the impact factor. The impact factor of a journal for a particular year is measured by the total number of citations received by the articles published in the journal during the preceding two years, divided by the total number of articles published in that journal during the preceding two years \cite{impactfactor}.
It is widely known and we shall also explain below that anomalous citations can purposefully manipulate impact factors. Since the Journal Impact Factor (JIF) is used to determine the importance of the contributions made by the journal in its own domain, it becomes essential to spot the anomalies in citation networks in order to understand if there is any misuse of citations to manipulate JIF and what are the reasons behind such misuse.

%, which can further be studied so as to ensure that the citation counts and thus the Journal Impact factor is not purposely colluded using anomalous citations.
Anomalous citations can be divided into two major categories:
\begin{enumerate}
    \item \textbf{Self-citations}:
    They refer to the citations made by  entities, i.e.,  journals or  authors, to themselves, which can, in certain cases, lead to a disproportionate increase in JIF of the entity  \cite{chakraborty2018good}.
    \item \textbf{Citation Stack}:
    It refers to the citation made by an entity, i.e., a journal or an author, to another entity, which can, in certain cases, be anomalously high and deviate from the usual behaviour of the entity  \cite{chakraborty2018good}.
\end{enumerate}

\citet{fowler2007does} showed that the more one cites oneself (self-citation), the more others cite them. \citet{ioannidis2015generalized} classified the phenomenon of self-citation to  different types -- direct, co-author, collaborative and coercive-induced self-citation. They further raised concerns over how inappropriate self-citations can affect impact factor, and suggested the usage of citation indices which are more foolproof to collusion using inappropriate self-citations. \citet{Bartneck2010DetectingHM} discussed how h-index can be inflated by authors by manipulating self-citation. \citet{SelfCitation}, in their study on analyzing the effect of excessive self-citation, argued for a new metric based on more transparency to help curb excessive self-citations, which may have a negative effect on our ability to truly analyze the impact and contributions of a scientific research in its research domain. However, limited work has been done in the field of analyzing citation stack.

In 2011, Thomson Reuters \cite{ThomsonReuter} (currently known as Clarivate Analytics) suspended 33 journals from its Journal Citation Report due to a very high rate of self-citations, which contributed to as much as 90\% of the JIF of those journals. Therefore, anomalous citation detection is an important task in bibliometrics.
%because it can give generate more insights for citation-stack and citation-cartel detection.

\textbf{Our Contributions}: In this paper, we study a citation network at the journal level, define the notion of anomalies for scientific citation networks, and detect the anomalous journal pairs with focus on the anomalies from one entity to another (citation stack), along with a confidence score indicating the correctness of our prediction. We also categorize them into different types, and figure out possible reasons for such anomalies.
 The major contributions of this paper are mentioned below:
\begin{itemize}
    \item We present a novel approach to detect the anomalies in a journal-level scientific citation network, and compare the results with the existing graph anomaly detection algorithms. 
%To the best of our knowledge, this is the first time work has been attempted on detecting anomalies in the citation network. 
%Our work focuses on detecting anomalous citations at the Journal level.
\item Due to the lack of ground-truth for citation anomalies, we introduce JoCAD, a novel dataset which consists of synthetically injected citation anomalies (Section \ref{Datasets_JoCAD}), and use it to evaluate our methodology and compare the performance of our method with the state-of-the-art graph anomaly detection methods. 
\item Our model is able to predict the anomalous citation pairs with 100\% precision and   86\% F1-score (Section \ref{experimental-results}). 

\item We further categorize the anomalies detected into various types and dig out possible reasons to develop a deeper understanding of the root causes. 

\item We also analyze our model on the Microsoft Academic Search dataset, a real-world citation dataset (Section \ref{Experiment_MAS}), labeled by human annotators (Section \ref{human-annotator}).

\item We interpret our results using a case study, wherein our results resemble the citations and Scimago Journal \& Country Rank (SJR) rating-change charts (Section \ref{case-study}). 

\item We further design an interactive web portal - `Journal Citations Analysis Tool' which can process any citation network and show journal-level anomalous citations. It helps the users analyze the temporal citation pattern of a given journal (Section \ref{portal}).

\end{itemize}

\noindent\fbox{%
    \parbox{0.78\columnwidth}{%
\textbf{Reproducibility:} The codes and the datasets are available at: \url{http://bit.ly/anomalydata}.}}

% \subsection{Terminology}

% \subsubsection{\textbf{Synchronous and Dianchronous Citations}}

% % We refer to the terms synchronous and dianchronous, defined by the authors of "A macro study of self-citation" \cite{macro}, and extend the definition of \textbf{synchoronous citations} to refer to the citations from a journal to another paper, and \textbf{dianchronous citations} to refer to the citations received by a journal.

% Extending the existing notions of synchronous and dianchronous citations\cite{macro}, we will use synchronous citations to refer to the citations from a journal to another journal, and dianchronous citations to refer to the citations received by a journal.

% \subsubsection{\textbf{Static and Dynamic Anomalies}}

% Following the usual definitions of static and dynamic settings, the anomalous pairs detected on the basis of the total number of citations between two journals over all the years are called \textbf{static anomalies}, and the pairs detected by year-wise analysis of citations are called \textbf{dynamic anomalies}.\\

% In this paper, we present methods to detect both types of anomalies using Box plot bucketing method and time series analysis.

\section{Related Work}\label{relatedwork}

The methodology used in this paper is closely related to the techniques used in anomaly detection in complex networks, and journal-level study of citation networks. Both these topics are well-studied; however, to the best of our knowledge, anomaly detection in journal-level citation network has not been studied yet.

\subsection{Anomaly Detection in Complex Networks}

%Anomalies or outliers refer to the data points or nodes, which do not conform to the expected behaviour. Anomaly detection algorithms are used to find the anomalies in a given set of data points.

Anomaly detection in networks is a widely explored area. \citet{Noble} proposed a subdue based method for anomaly detection in graphs. 
\citet{Hodge2004} wrote a survey of machine learning and statistical techniques for outlier detection, and categorized the outlier detection algorithms into three major types, namely, unsupervised clustering, supervised classification and semi-supervised recognition. 
The anomalies themselves have been categorized into two broad types \cite{chen} -- \textit{white crow} anomalies, which can be easily identified as outliers, and \textit{in-disguise} anomalies, which try to hide themselves and are thus difficult to detect.  \citet{moon} presented OutRank, an algorithm to detect white-crow anomalies by assigning a similarity score to objects. \citet{Eberle2007MiningFS} detected in-disguise anomalies, which they called ``structural anomalies".
We present two methods that cover both  kinds of anomalies mentioned above. The box plot method in Section \ref{sec:method} detects white-crow anomalies as it identifies high number of citations between journals (which is an easily noticeable feature) after bucketing them according to their sizes. The time-series method in Section \ref{sec:method} detects in-disguise anomalies as it analyzes citations over the years and detects outliers from the usual behaviour (which can not be noticed from an overall study of the data).

Anomalous groups in a citation network can be seen as communities with high citation activity with each other.
\citet{Sun2005NeighborhoodFA} presented algorithms for community detection in bipartite graphs. They introduced two kinds of functions for the purpose of community detection and outlier detection.
% Research work has been done in community detection in complex graphs - Girvan and Newman \cite{Girvan7821} used the idea of centrality and betweenness for detecting communities in graphs with known community structures as well as unknown community structures. Chakrabarti \cite{Chakrabarti} presented algorithms to find such communities and identified outliers by computing distances between these comunities.
%
Oddball algorithm was proposed for unsupervised anomaly detection in weighted graphs, wherein each node is treated as an ego, and the neighbourhood network around it is studied as the egonet \cite{oddball}.
\citet{Noble} detected anomalies in graph-based data by categorizing them in two ways - anomalous sub-structures and anomalous sub-graphs. They claimed that these anomalous sub-structures and sub-graphs are extremely rare as compared to ``usual" sub-structures and sub-graphs.
%This claim coincides with our model's results. The percentage of pairs of journals found anomalous in the MAS data set (Section \ref{datasets}) is very little (only 3.9\%) (Section \ref{experimental-results}). 
 \citet{Chandola:2009:ADS:1541880.1541882} presented a survey which provided classified anomalies into three types:
 
\begin{itemize}
    \item \textbf{Point Anomalies:} 
     It refers to the individual data points which are anomalous with respect to the entire data.
     For example, consider the real life scenario of the price of a stationary item. The cost of the item suddenly rises and becomes twice the previous cost. The new cost is thus, an example of a point anomaly.
    \item \textbf{Contextual Anomalies:} 
     If a data instance is anomalous only in a particular context, then it is a contextual anomaly. It is also known as a conditional anomaly.  \citet{Huber} designed a method to process event streams from technical systems for timely detecting anomalies and thus helping prevent huge industrial losses.
    \item \textbf{Collective Anomalies:}
     If the existence of certain data points individually is not anomalous but their collective existence collectively is, then they are called collective anomalies.
 \end{itemize}
In our study, we focus on the collective anomalies.

The standard anomaly detection algorithms for graphs could not be used for detecting pairwise anomalies in a citation network because the subtle signal strengths of the anomalies at the journal level in the citation network could not be detected by the pre-existing graph algorithms. An alternate way of modelling this problem can be using supervised machine learning techniques, which can be used to learn the suitable feature representation that can help in detecting anomalous behaviour among journals, authors, paper, etc. However, due to the lack of a ground-truth dataset of considerable size which has sufficient examples of anomalous behaviours of the above entities, machine learning algorithms could not be applied. We empirically show that existing graph anomaly detection methods could not work on the problem under consideration (Section \ref{experimental-results}).
%The results can be found in the Experiments section of the paper. 
This motivated us to study detection of pair-wise anomalies.

\subsection{Journal-level Study of Citation Networks}

\citet{Hummon1989ConnectivityIA} modelled the connectivity in citation networks as an exhaustive search problem and applied a depth-first-search variant to quantify the similarity in the network. They focused on the similarity of edges. \citet{Zhang} presented a clustering algorithm which calculated the similarity between adjacent nodes and formed clusters based on that score. 

Study of citation network is a sufficiently explored topic. It dates back to 1965, when \citet{solla} studied the citation network and made inferences about the nature of references of a scientific paper with respect to its field and length. Various facets of citation networks have been studied, most widely explored being the network of papers and its outgoing citations. \citet{Osareh} studied citation network and its applications. \citet{Zhao:2008:CPC:1458082.1458125} studied the evolution of heterogeneous citation networks.

Researches have also focused on the citation interactions among  authors using citation analysis. \citet{Prabha} carried out a pilot study on the reasons behind the citations in scientific papers. He found that many of the citations made by papers are not essential for the results produced by it. \citet{Cronin2002} studied paper reference patterns with respect to the authors of the paper. 
They defined an author's \textit{citation identity} as the authors they cite in their work, and \textit{citation image} as the authors who cite them. They argued that the reference styles of authors are a form of watermark for their work.

% There can be other important aspects of a publication that could be considered for research, including: the geographic location of the publication ("Citation behavior: Classification, utility, and location" by V.Cano \cite{cano}), and the venue of publication ("Toward alternative measures for ranking venues: a case of database research community" by Yan and Lee \cite{Yan:2007:TAM:1255175.1255221}).

One important application of  anomaly detection in citation network is to find the possibility of collusion among entities.   \citet{cartel} came up with an algorithm to detect citation cartels \cite{journal}. It was a community detection algorithm focusing on anomalous pair of citations. The authors also discussed the possibility of some anomalous pairs being cases of citation cartels. Claiming whether a citation is a result of possible collusion or not is a non-trivial task, and requires an in-depth examination of every citation. In this paper, we focus only on the anomalies in the citation graph, and we do not try to claim any intentions behind those anomalous citations. 

\section{Datasets}\label{datasets}
For the purpose of this study, we consider two different datasets. One of the major challenges while devising the experiments was the lack of ground-truth for anomalous citations in the real-world dataset. We thus introduce a journal citation anomaly dataset (called JoCAD), a dataset which consists of synthetically injected anomalies (see Section \ref{Datasets_JoCAD}). We use this for the purpose of testing our model and reporting the final accuracy in terms of the F1-score  (Section \ref{experimental-results}). We also use the Microsoft Academic Search Dataset, a real-world citation dataset, for carrying out a case-study on the anomalous pair of journals detected by our model (Section \ref{case-study}).

\subsection{Microsoft Academic Search Dataset}\label{Datasets_MAS}
The Microsoft Academic Search (MAS) Dataset \cite{chakraborty2018good,MAS} consists of 1.6 million publications and 1 million authors related to the Computer Science domain. It has various metadata information about a paper, including the year of publication, keywords of the paper, the references of the paper, the specific field (AI, algorithms, etc.) the paper is targeted towards, and the abstract of the paper. The dataset spans from 1970 to 2014. This dataset has been extensively used in the past for bibliographic analysis \cite{chakraborty2015categorization,chakraborty2018universal}.

We preprocessed the dataset according to the needs of our model. We removed the metadata which were not useful for the study. These included keywords, abstract, and the related fields of the paper. Table \ref{tab:MAS} shows statistics of the dataset.

\begin{table}[!h]

\begin{center}
\caption{ Microsoft Academic Search Dataset \cite{chakraborty2018good,MAS}}\label{tab:MAS}
\begin{tabular}{ c | c } 
 \hline
 Entity & Number \\
 \hline 
 No. of papers & 1.6 million  \\
 No. of citations & 6 million \\
 No. of journals & 1,800 \\
 No. of journal pairs & 8,500\\
 \hline
\end{tabular}
\end{center}
\vspace{-5mm}
\end{table}

% \begin{table}[h!]

% \begin{center}
% \caption{Statistics related to the  Dataset}\label{tab:MAS}
% \begin{tabular}{ c c } 
%  \hline
%  Entity & Number \\
%  \hline 
%  Papers & 1.6 Million  \\
%  Citations & 6 Million \\
%  Journals & 1.8 Thousand \\
%  Journal pairs & 8.5 Thousand \\
%  \hline
% \end{tabular}
% \end{center}
% \end{table}

\subsection{JoCAD - Our Synthetically Injected Anomaly Dataset}\label{Datasets_JoCAD}
For anomaly detection in citation graphs, the ground-truth is not well-defined. Hence, we created a synthetic dataset, wherein we synthetically injected anomalies to serve as the ground truth in our further experiments. The synthetic data generation process is motivated by \citet{Hayat2017}.

Our synthetic dataset contains 100 journals, each having a journal index  between 0 and 99, and containing citation information for the journals ranging over 20 years (2000-2020).

We assign the number of papers in the journal as a random number in a pre-decided range. Similarly, we assign the citation count between each pair of journals as a random number in a pre-decided range which we consider as the normal citation behaviour of the journal pairs.

For injecting anomalies, we intuitively list properties of an anomalous pair of journals. The properties consist of a sudden spike in the number of citations from one journal to another or from both journals to each other in some random year; or a gradual increase from one journal to another or from both journals to each other, which result in an anomalously high number of citations over all the years. Using these methods with reasonable randomness, we create a total of 110 anomalies. The final dataset consists of 110 anomalous pairs out of total 10,000 pairs of journals. 

The types of anomalies we inject into the dataset can be broadly classified into five types:
\begin{enumerate}
    \item If the number of citations from  journal $J_i$ to journal $J_j$ in a particular year is significantly more than the normal citation behaviour of the journal pairs, i.e., there is a sudden spike in the number of citations from $J_i$ to $J_j$ in that year, then the citation from $J_i$ to $J_j$ is an instance of {\bf Type 1 anomaly}.
    \item If the number of citations from $J_i$ to  $J_j$ increases significantly over successive years, with the difference in the number of citations between any two years $Y_i$ and $Y_j$ being greater than the expected difference in the number of citations between any journal pair, then the citation from $J_i$ to $J_j$ is an instance of {\bf Type 2 anomaly}.
    \item If the number of citations from $J_i$ to $J_j$ in the year $Y_k$ is significantly higher than the average number of citations between any journal pairs, and the number of citations from $J_j$ to $J_i$ in the year $Y_{k+1}$ is significantly higher than the average number of citations between any journal pairs, then both the citations from $J_i$ to $J_j$ and vice-versa are instances of {\bf Type 3 anomaly}.
    \item If the number of citations from $J_i$ to  $J_j$ in the year $Y_i$ is greater than or equal to double the number of citations received by $J_i$ from $J_j$ in the previous year $Y(i-1)$, then the citation from $J_i$ to $J_j$ is an instance of {\bf Type 4 anomaly}.
    \item If the number of citations from $J_i$ to  $J_j$ in the year $Y_i$ is greater than or equal to double the number of citations from $J_i$ to $J_j$ in the previous year $Y_{i-1}$, then the citation from $J_i$ to $J_j$ is an instance of {\bf Type 5 anomaly}.
\end{enumerate}

\section{Methodology}\label{methodology}
In this section, we first introduce the problem definition, followed by the proposed methodology.

\begin{figure*}[!]
    \centering
    \includegraphics[width=15 cm]{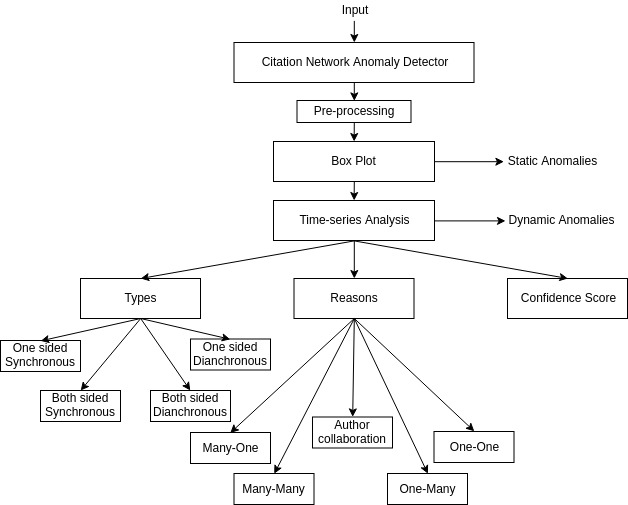}
    \caption{Flow diagram of our methodology.}
    \label{fig:methodology}
    %\vspace{-5mm}
\end{figure*}

\subsection{Problem Definition}
Given a set of papers $P$, where each paper has 4 attributes -- journal name, authors, citations, and the year of publication, as the input to the system, the task is to predict the journal-to-journal anomalies with a confidence score of the prediction made,  classify the anomaly into its type and suggest possible reasons for the existence of the same.

\subsection{Proposed Methodology}\label{sec:method}
We use a two-fold approach (as shown in Figure \ref{fig:methodology}) to classify the citation anomalies --- box plot and time-series analysis.

\subsubsection{Box Plot Bucket Analysis:}

We start by clustering the journals into buckets based on similar behaviour in terms of the total number of papers published in the journal.
For each journal in our database, we define the usual behaviour of the given journal by analyzing the behaviour of all the journals lying in the same bucket as the given journal, i.e., all the journals having similar number of papers as that of the given journal, and then use the box plot method to predict the anomalies.

For each possible bucket, we pair each of the papers in the current bucket with each of the papers in the bucket of the given journal. We refer to the set of all these paper-paper pairs as a grid. We now detect the anomalies using box plot. To calculate the outlier behaviour, using a box plot, we calculate the first quartile and the third quartile\footnote{
The first quartile is the 25$th$ percentile of the data and the third quartile is the 75$th$ percentile of the data. The second quartile refers to the median of the distribution.} of the data points corresponding to the journal-journal citation counts in the selected grid. 

We define anomalous journal-journal pairs as the data points which lie outside the area:
[(first-quartile - ($1.5\times IQR$)) , (third-quartile + ($1.5\times IQR$))], where interquartile range $IQR$ = (third-quartile - first-quartile).

The anomalies detected by the box plot method are static anomalies as these anomalies are based on the total number of citations between two journals over all the years.

\subsubsection{Time-series Analysis:}

Once we have the pairs declared anomalous by the box plot, we analyze each of them with time-series analysis to confirm if there exists some year during which there was a sudden change in their usual citation behaviour. 
The anomalies detected by the time-series analysis method are dynamic anomalies as these anomalies are susceptible to change every year.

To define the ``usual behaviour", we use the empirical rule. The empirical rule, also called the ``68-95-99.7 rule'' \cite{pukelsheim1994three}, is a statistical rule that states the percentage of values of a Gaussian distribution that \textcolor{black}{lies} within 2, 4, and 6 standard deviations around the mean. According to the rule, 2 standard deviations around the mean cover 68.27\% of the values, 4 standard deviations around the mean cover 95.45\% of the values, and 6 standard deviations around the mean cover 99.73\% of the values.

To check the usual behaviour for a pair of journals for a given year (say $Y$), we list the number of citations from one journal to another till $Y$, and calculate mean and standard deviation from the distribution. Then, as stated by the empirical formula, 99.73\% of the data is expected to lie within 6 standard deviations around the mean. Hence, if the number of citations from one journal to another exceeds this range, it is suggested to be anomalous by the time-series analysis. 

To understand the different types of unusual behaviour, we extend the existing notions of synchronous and dianchronous citations \cite{macro} and use synchronous citations to refer to the citations from a journal to another journal, and dianchronous citations to refer to the citations received by a journal.

For a given pair of journals, we check four types of behaviours for each year:

\begin{enumerate}
    \item \textbf{One sided synchronous citations}:
    This consists of detecting the unusual behaviour of outgoing citations from one journal to the other for each side, i.e., if the two journals are $J_1$ and $J_2$, the above test is performed separately for the outgoing citations from $J_1$ to $J_2$ and for the outgoing citations from $J_2$ to $J_1$, for each year as stated above.
    \item \textbf{One sided dianchronous citations}:
    This consists of detecting the unusual behaviour of incoming citations from one journal to the other for each side i.e., if two journals are $J_1$ and $J_2$, the above test is performed separately for the incoming citations from $J_1$ to $J_2$ and the incoming citations from $J_2$ to $J_1$, for each year as stated above.
    \item \textbf{Double sided synchronous citations}:
    This consists of detecting the unusual behaviour of outgoing citations from one journal to the other for both the journals simultaneously, and if both of the sides lie in the outlier range, this pair is said to have double sided synchronous behaviour.
    \item \textbf{Double sided dianchronous citations}:
    This consists of detecting the unusual behaviour of incoming citations from one journal to the other for both the journals simultaneously, and if both of the sides lie in the outlier range, this pair is said to have double sided synchronous behaviour.
\end{enumerate} 

Checking both synchronous and dianchronous citation behaviours is essential because a particular number of citations from one journal to another can be in the normal range for the citing journal but not in the normal range for the cited journal, or vice-versa. This kind of case can happen in many instances. For example, one unpopular journal suddenly  increases citations to a very popular journal and to an unusually high level; or one very popular journal suddenly increases citations to an unpopular journal. In both cases, only one side of the single side anomalies will be evident from the distribution based check described above.

\subsubsection{Reasons for Possible Anomalies:}
We categorize the detected anomalies hinting at five possible reasons for their occurrence, namely, many-many anomaly, many-one anomaly, one-many anomaly, one-one anomaly, and previous-author collaboration.
For the first four, we supplement a metric that captures the essence of the citation graph being highly crowded at both the ends (journals $J_1$ and $J_2$) for many-many anomaly, and so on.

To do this, we define the ``normal behaviour" of the citations {\em given by a paper to a journal} by first creating a Gaussian distribution with the citation count from a paper to a journal as the data points and then, using the empirical formula, we use $\mu \pm \sigma$ to define the normal behaviour of nearly 70\% of the data points. 
Now consider all the papers in the citing (sender) journal, i.e., the journal $J_1$, and count the number of papers in the journal which contribute to $J_2$ (cited/receiver journal) that is at least the normal behaviour: $paper-to-journal\ count > \mu \pm \sigma$. Dividing this by the total number of papers in published in $J_1$, we get the percentage of papers (vertices in the citation network) in the $J_1$'s side which is crowded in terms of the citation count they produce for $J_2$, denoted as `sender-percentage'.

Now, we define the ``normal behaviour" of the citations received by a paper from a journal, by creating a Gaussian distribution with the citation count from a journal to a paper as the data points, and then, we use the empirical formula to define the normal behaviour of the data points. 
Similar to the previous case, based on the above definition of the normal behaviour, we now consider all the papers in the receiver journal $J_2$, and count the number of papers in $J_2$ which are cited by $J_1$ (the sender journal) more than or equal to the normal behaviour: $journal-to-paper\ count > \mu \pm \sigma$. Dividing this by the total number of papers in $J_2$, we get the percentage of papers (vertices) in the receiver's side which is crowded in terms of the citation count they receive from $J_1$, denoted as `receiver-percentage'.

Now we are ready to categorize an anomaly as many-many, many-one, one-many, or one-one based on the following criteria:
\begin{enumerate}
    \item \textbf{Many-many Anomaly:}
    If the sender-percentage is $>75\%$ and the receiver-percentage is $>75\%$, then it is the case of a many-many anomaly since both the receiver's and the sender's sides are extremely crowded.
    \item \textbf{Many-one Anomaly:}
    If the sender-percentage is $>75\%$ and the receiver-percentage is $<25\%$, then it is the case of a many-one anomaly since the receiver's side is not crowded while the sender's side is extremely crowded.
    \item \textbf{One-many Anomaly:}
    If the sender-percentage is $<25\%$ and the receiver-percentage is $>75\%$, then it is the case of a one-many anomaly since the sender's side is not crowded while the receiver's side is extremely crowded.
    \item \textbf{One-one Anomaly:}
    If the sender-percentage is $<25\%$ and the receiver-percentage is $<25\%$, then it is the case of a one-one anomaly since both the sender's and receiver's side are not crowded.
\end{enumerate}

\begin{figure*}[!t]
    \centering
    \includegraphics[width=1.1\columnwidth]{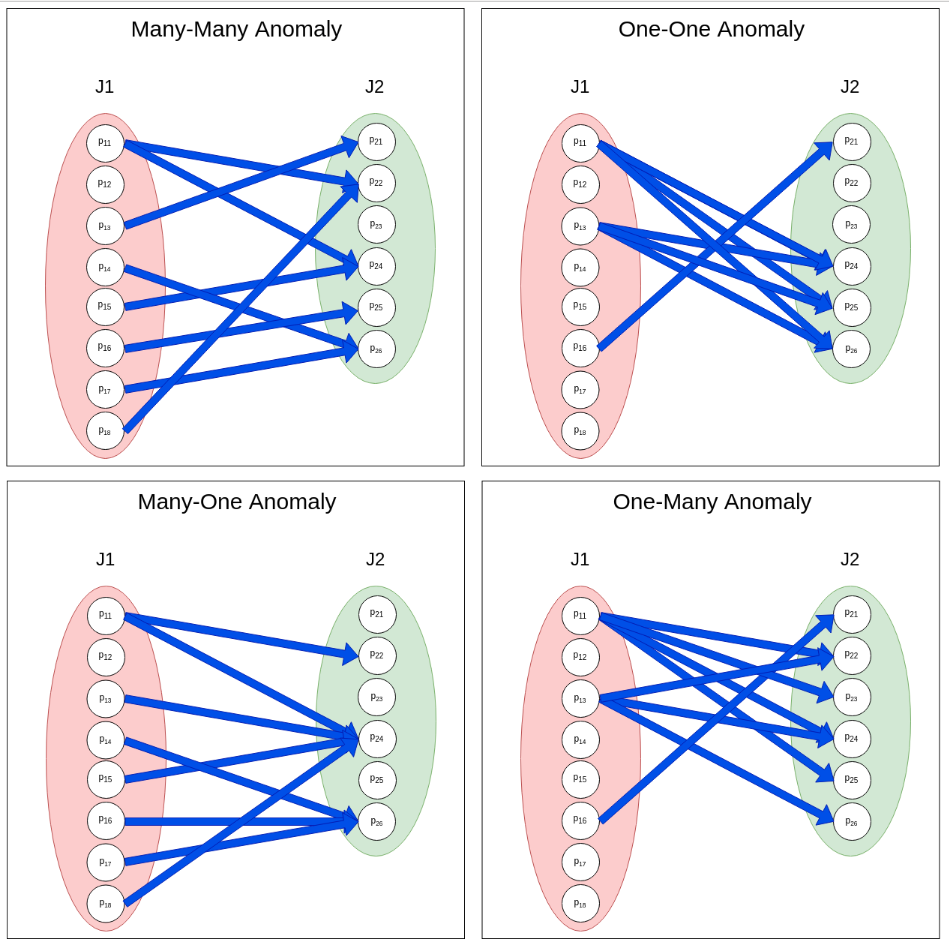}
    \caption{Pictorial representation of four of the five possible reasons of anomalies. The fifth reason (effect of collaboration) is not shown here.}
    \label{fig:reasons}
\end{figure*}

Figure \ref{fig:reasons} diagrammatically shows the four reasons listed above.

To explain the reasons for the anomalies, we also look into the previous collaboration of the authors of a publication and count the number of such occurrences between the journal-journal pairs.

In particular, we detect if at the time of publishing a paper in year $Y$ any of the authors of the paper has previously collaborated with any of the authors of the paper it is citing. We then increment a counter by $1$. We do so for paper-pairs consisting of every paper in the sender-journal $J_1$ and all the papers it cites in the receiver-journal $J_2$, and thus calculate the total number of \textbf{previous-author-collaborations}.

\subsection{Confidence Score}\label{confidence}

Our model finally assigns a confidence score to every anomaly detected, which states how confident the model is on the pair of journals being anomalous in a year \cite{ConfScore}. A confidence score is first assigned by box plot method to each anomaly, then by the time-series method, and at last both the scores are combined, giving the final confidence score. 
We use the $tanh$ function\footnote{$tanh$ is a hyperbolic function which gives a value ranging from -1 to 1, and in case of only positive inputs, 0 to 1.} for calculating the individual confidence scores from the two methods -- box plot and time-series.  

One intuition is that if our method states a pair to be anomalous, we are at least $50\%$ sure of it being anomalous. Following this intuition, we scale the confidence scores given by individual methods to be from $0.5-1$ instead of $0-1$. Another intuition is that in the time-series method, both-sided anomaly is a more unusual event than a single-sided anomaly. This is because a both-sided anomaly in a particular year means both the journals went beyond the normal behaviour and cited the other. We state the confidence of a both-sided anomaly to be at least $75\%$. Therefore, a both-sided anomaly is scaled between $0.75-1$ instead of $0-1$. 

\begin{figure*}
    \centering
    \includegraphics[width=15 cm]{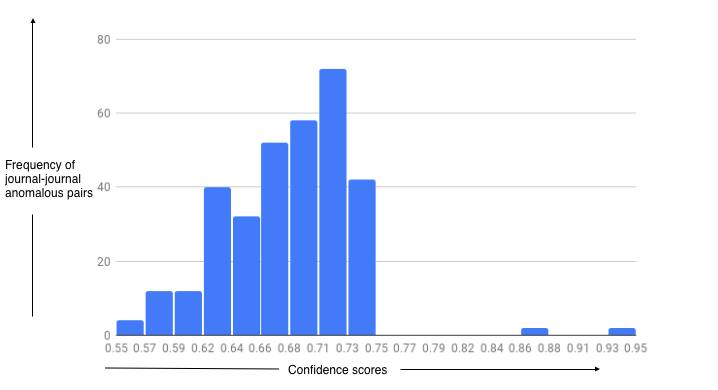}
    \caption{Histogram of confidence scores of anomalies in the MAS dataset.}
    \label{fig:histogram}
\end{figure*}

For combining the scores, we choose to take the average of the two scores. This is done because both the methods have equal importance in the detection of the anomaly. 
Figure \ref{fig:histogram} shows the histogram of the confidence scores of anomalies detected from the MAS dataset. We analyze the most anomalous pair with confidence score 0.93 and find  that it is logical for the pair to be anomalous (see case studies in Section \ref{case-study}).

\section{Experiments}\label{experiments}

\subsection{Comparison of Time-series Analysis with SJR ratings}\label{experimentA}

To evaluate the performance of the time-series method for anomalous behaviour detection, we use the SJR ratings to check changes in ratings of journals stated anomalous in some particular years. Scimago Journal \& Country Rank, better known as SJR, is a journal rating system  \cite{SJR}. It uses data similar to what is used by H-index, and gives year-wise ranks to journals. It is a well-recognised rating system for journals and conferences.

We scraped the SJR ratings over the years for all the journals present in the MAS dataset  from the portal made for SJR ratings. The comparison of sudden increase in citations to each other for pairs of journals mostly resonates with the respective SJR rating changes of the citation-receiving journal in the next two years. The case-study of the most anomalous pair of journals also supports the same result (see Section \ref{case-study}).

\subsection{Experimental Results}\label{experimental-results}

For anomaly detection in citation graphs, there is no well-defined ground-truth or a proper evaluation metric. Hence, for the purpose of evaluating the performance of our method, we created a synthetic dataset wherein we synthetically injected anomalies~\ref{datasets}. They were treated as the ground-truth in our experiments, and we report our results based on them. We use precision, recall and F1 score for the final evaluation. 
%F1 score is the weighted harmonic mean of Precision and Recall. F1 score is the correct metric for anomaly detection because both precision and recall are equally important. Precision is the ratio of actually anomalous pairs with the pairs stated anomalous by the model. Recall is the ratio of actually anomalous pairs with all the anomalous pairs. 

%The general metric used in any statistical study- Accuracy is not the correct metric for the evaluation for our model. Accuracy is the ratio of actually anomalous pairs and actual non-anomalous pairs with all the pairs in the test set. It is not the correct metric for anomaly detection because anomalies are very rare in the dataset, and hence, the number of true negatives is always very large. The large number of true negatives dominate the accuracy score, and true positives do not get get enough weight in this metric. 

\textbf{Baseline Methods:} Due to the lack of any existing baseline for the given task, we decided to use the existing graph anomaly detection algorithms and each of the methods used by us, namely, the box plot method and the time-series method, to act as baselines for evaluating our proposed model.

We used K-means clustering, one of the popular graph anomaly detection algorithms, but it could only detect 38 anomalous pairs out of the total of 110. The detected pairs are the pairs with a large number of citations on both sides i.e., to and from one journal to another over the years. These are clearly visible by taking the summation of the citations on both sides, similar to the box plot method. However, K-means clustering algorithm is unable to detect other types of anomalous pairs, and thus has a lower recall and F1 score as compared to our method.

\begin{table}[h!]
\label{F1_scores}
\begin{center}
\caption{Results of anomaly detection methods and our models.} \label{tab:result}
\begin{tabular}{ c c c c } 
 \hline
 Method  & Precision (\%) & Recall (\%) & F1 Score (\%) \\
 \hline 
 K-means & 100 & 34.5 & 67.2 \\
 Box plot & 59.19 & 93.63 & 72.53 \\
 Time-series & 86.73 & 77.27 & 81.73 \\
 Our model & 100 & 75.45 & 86.01 \\
 \hline
\end{tabular}
\end{center}
\end{table}

Table \ref{tab:result} shows the comparative results of the baselines and our model. The F1 score of our model comes out to be $86.01\%$ which is higher than that of the baselines. Also, the precision of our model comes out to be $100\%$, which means that all the pairs stated anomalous with the agreement of both the methods are actually anomalous.

\subsection{Results on the MAS Dataset}\label{Experiment_MAS}

We ran the model on the real-world MAS dataset. The histogram of confidence scores of the anomalies found in the dataset is shown in Figure \ref{fig:histogram} (explained in Section \ref{confidence}). Out of total 8.5 thousand journal-pairs, total of 328 anomalies are found. The number of unique pairs of anomalous journals (irrespective of the year of anomaly) is 230, and the total number of journals found in any anomaly is 103. We observe that the occurrence of anomalies is rare (only $3\%$), and the occurrence of both-sided anomalies is extremely rare (only $4$ cases out of $8.5$ thousand pairs), as shown in Table \ref{tab:distribution}

\begin{table}[h!]
\begin{center}
\caption{Distribution of types of journal-pair anomalies found in the MAS dataset.}\label{tab:distribution}
\begin{tabular}{ c c c } 
 \hline
 Type & Number & Percentage of total pairs \\
 \hline 
 Single (One sided) & 324 & 3.85\% \\
 Double (Both sided) & 4 & 0.04\%\\
 Total & 328 & 3.90\% \\
 \hline
\end{tabular}
\end{center}
\end{table}

\begin{figure}
    \centering
    \includegraphics[width=\columnwidth]{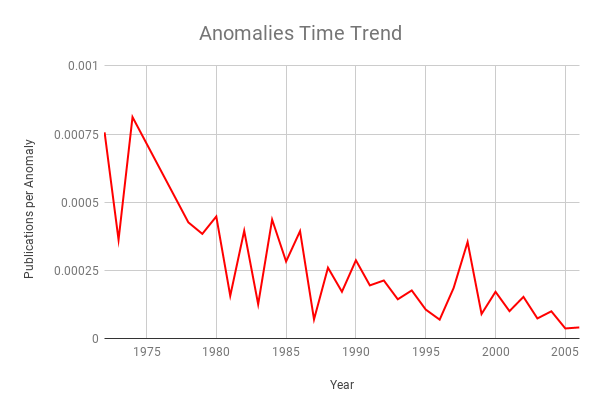}
    \caption{Average anomalous publications throughout the years.}
    \label{fig:year}
\end{figure}

We plot the average number of publications per year that are anomalous, i.e., the ratio of the number of anomalous publications in a year to the total number of publications in that year. The temporal frequency mapping of anomalies is shown in Figure \ref{fig:year}. We can infer that the frequency of the relative year-wise anomalous journals  decreases over the years. This decrease might be due to a huge boost given to the academic research in the field of Computer Science, which led to an increase in the total number of journals. This might have decreased the ratio over time. Another reason could be more stringent rules and regulations imposed in the scientific community, which could have led to a decrease in anomalous activity amongst journals. 

\begin{figure}
    \centering
    \includegraphics[width=\columnwidth]{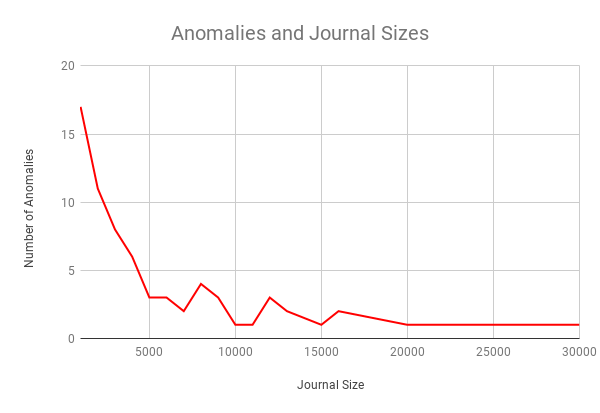}
    \caption{Fraction of anomalous papers published in a journal with the size of the journal. }
    \label{fig:size}
\end{figure}

Next, we plot the trends with respect to the size of the anomalous journals, defined in terms of the number of papers published in it. Figure \ref{fig:size} shows the decline of anomalous activities as the size of a journal increases. This might be due to a bigger journal having a wider variety of papers, research domains, and authors. Another reason might be the journals which have gained recognition and prestige over time are likely to receive high number of novel and high-quality papers which should be published. Thus, prestigious and renowned journals may have a bigger journal size, which are less likely to be anomalous.

\begin{figure*}[t!]
    \centering
    \includegraphics[width=15 cm]{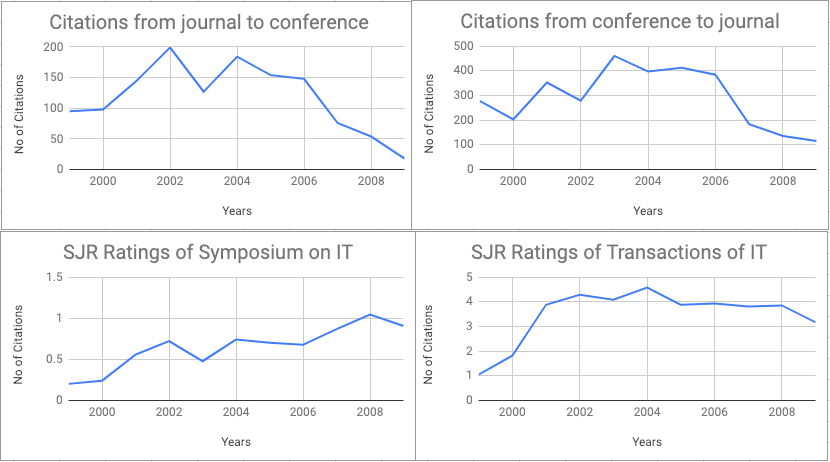}
    \caption{The number of citations from Symposium on Information theory to Transactions of Information Theory over years.}
    \label{fig:case}
\end{figure*}

\subsection{Human Annotators}\label{human-annotator}
Two human annotators\footnote{The annotators were experts in data mining domain and their age ranges between 25-35 years.} were asked to independently label the anomalies detected by our method on the MAS dataset as anomalous or non-anomalous. The annotation was done on the basis of the number of citations between the journals over years, and relevant metrics such as the average number of citations over all the years between the journals and the popularity of both journals. They checked the popularity of journals based on factors like impact factor and cite score.  

Each annotator annotated all 328 anomalous pairs. They placed 38 pairs and 25 pairs respectively in the non-anomalous category. The inter-annotator agreement was $0.78$, based on \textit{Cohen's kappa coefficient}.

\subsection{Case Study}\label{case-study}

We conduct a case study for the anomalous pair with the highest confidence in the MAS dataset.
The pair is the journal \textit{IEEE Transactions on Information Theory} and the conference \textit{IEEE International Symposium on Information Theory}, and the confidence score of being an  anomaly is $93\%$. IEEE has a professional society, namely \textit{The Information Theory Society} \cite{itsoc}, under whose umbrella lies its main journal - \textit{IEEE Transactions on Information Theory} (IEEE TIT) and its flagship conference, \textit{IEEE International Symposium on Information Theory}. The symposium is a conference for researchers to meet and discuss work done in the field of information theory. The work can be previously published or unpublished. 

As the work can be previously published, and the conference is a part of the same society as the journal, many authors of the published papers in the journal could attend the conference to display their work. Hence, it is natural for the two to have numerous citations to each other over the years. Because of this, there can be two reasons for a citation:

\begin{enumerate}
    \item A paper has already been published in the journal and the authors were invited to the conference to present and discuss their work.
    \item A paper was not accepted by the journal and the authors presented their work \textcolor{black}{at} the conference. Later, some other researcher used their work and cited them. The paper then got published in the journal.
\end{enumerate}

Some citations may not exist because of the journal and the conference being a part of the same society. There would be many instances where a paper not published in the journal but presented at the conference had cited a paper previously published in the journal.

\begin{figure*}[!t]
    \centering
    \includegraphics[width=15 cm]{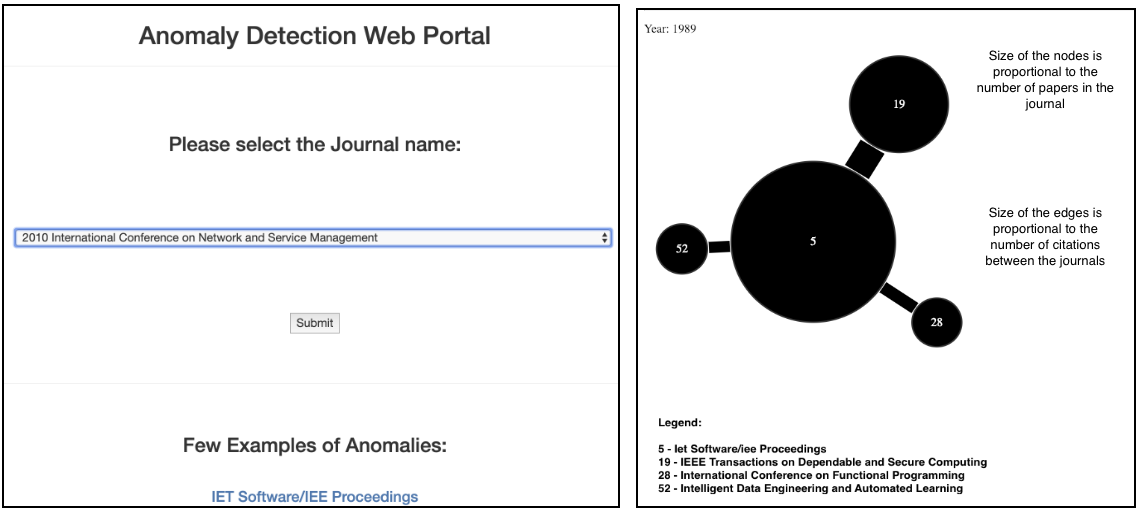}
    \caption{A snapshot of our portal. It shows the result of detecting multiple anomalous relationships for the journal IET software/IEEE Proceedings in the year 1989.}
    \label{fig:Portal_2}
\end{figure*}

Charts in Figure \ref{fig:case} show the number of citations from Symposium on Information Theory to IEEE TIT, the change in SJR rating of IEEE TIT over the years, the number of citations from IEEE TIT to Symposium on Information Theory, and the change in SJR ratings of Symposium on Information Theory over the years.

A clear resemblance in citations and SJR rating-change charts \cite{scimagojr.com} for both the journal and the conference shows that the findings of time-series analyzes correspond to the changes in the SJR rating of IEEE TIT. As stated in the introduction, JIF depends upon the number of citations the journal had received in the two previous years. A sudden spike can be seen in the citations from the conference to the journal \textcolor{black}{in the year} 2003. At the same time, a spike can be seen in the SJR rating of the journal \textcolor{black}{in the year} 2004, which shows that the increased citation count has contributed towards \textcolor{black}{an} increase in the rating of the journal.

\section{A Web Portal}\label{portal}
We use the journal level citation network created using the MAS dataset to develop a portal which allows the user to find the anomalies corresponding to a journal for all the possible years. It provides an interactive graphical interface, which visually depicts the anomalies in a citation network.
The user can enter the details of the journal, based on which the portal provide the user with a year-wise analysis of the anomalies. A screenshot of the portal is presented in Figure \ref{fig:Portal_2}.

For all pairs of journals in our dataset which form an anomalous pair with the journal queried by the user, we display the year-wise anomalies in an anomaly graph. This graph depicts each of the journals as nodes, while the citations between them is represented by the edges. The size of a node is proportional to the number of papers in the journal, i.e., a larger size of the node indicates that the journal has published a higher number of papers. The size of an edge depicts the number of citations between the two journals, i.e., a thicker edge between two journals indicates a higher number of citations between them. The number on top of each node represents its journal index, which is mapped to the respective journal name in our dataset.\smallskip{}

\noindent\fbox{%
    \parbox{0.98\columnwidth}{%
A beta version of the portal is live and can be accessed here: \url{https://journalcitationsanalysistool.herokuapp.com/}.}}

\section{Conclusion}\label{conclusions}
In this paper, we presented a novel model to detect the journal-level anomalous pairs. The model also gives the confidence score by analyzing both static and dynamic anomalies using box plot bucketing method and time-series analysis. We also curated JoCAD, a novel dataset, which consists of synthetically injected citation anomalies. We further ran our model on two datasets, namely, JoCAD and the Microsoft Academic Dataset, and used it to evaluate our methodology. We achieved 86\% F1 score on JoCAD in the comparison of our method with the standard graph anomaly detection methods.

We interpreted our results in a case study using the Microsoft Academic Dataset, wherein our results resemble the citations and Scimago Journal \& Country Rank (SJR) rating-change charts. We experimentally showed a high similarity between the time-series trends predicted by our method and the time-series trends as reflected by the SJR. We further designed an interactive web portal - `Journal Citations Analysis Tool` which given the citation network as an input, shows the journal-level anomalous citations, and helps users analyze the temporal citation pattern of a given journal.

Future work in this direction can include an in-depth study about the reasons for the anomaly detected as well as extend the work in developing methods for the detection of citation cartels. We acknowledge that while curating the dataset, JoCAD, we could have possibly introduced a human bias in it, which can be reduced in the future. We would like to explore the anomalous citation behavior in other domains as well. Other relevant future work can take into account the textual context of a citation made, and calculate its similarity with the content of the paper being cited. This will help us understand the relevance of the citation made and thus help predict anomalous citations better. 
\section*{Acknowldgement}
T. Chakraborty would like to acknowledge the support of Ramanujan Fellowship, DST (ECR/2017/00l691), and the Infosys Centre of AI, IIIT-Delhi, India. 

\bibliographystyle{ACM-Reference-Format}  \bibliography{references.bib}

\end{document}